\documentclass[runningheads]{llncs}

\usepackage{graphicx}
\usepackage{enumitem}
\usepackage[caption=false,font=footnotesize]{subfig}
\usepackage{multirow}
\usepackage{amsmath,tabularx}
\usepackage{breqn}
\usepackage{float}
\usepackage{array}
\newcolumntype{L}[1]{>{\raggedright\let\newline\\\arraybackslash\hspace{0pt}}m{#1}}
\newcolumntype{C}[1]{>{\centering\let\newline\\\arraybackslash\hspace{0pt}}m{#1}}
\newcolumntype{R}[1]{>{\raggedleft\let\newline\\\arraybackslash\hspace{0pt}}m{#1}}
\usepackage{ulem}

\usepackage[section]{placeins}

\begin{document}

\title{Subjective Metrics-based Cloud Market Performance Prediction\thanks{This is a post-peer-review, pre-copyedit version of an article published in the 21st International Conference on Web Information Systems Engineering (WISE 2020).}}
%
%\titlerunning{Abbreviated paper title}
% If the paper title is too long for the running head, you can set
% an abbreviated paper title here
%
%\titlerunning{Abbreviated paper title}
% If the paper title is too long for the running head, you can set
% an abbreviated paper title here
%
\author{Ahmed Alharbi\and
Hai Dong\thanks{Corresponding author} }
\authorrunning{A. Alharbi and H. Dong}
% First names are abbreviated in the running head.
% If there are more than two authors, 'et al.' is used.
%
\institute{School of Science, RMIT University, Melbourne, Australia \\
\email{\{ahmed.alharbi,hai.dong\}@rmit.edu.au} \\
}

\maketitle              % typeset the header of the contribution
\begin{abstract}
This paper explores an effective machine learning approach to predict cloud market performance for cloud consumers,  providers and investors based on social media. We identified a set of comprehensive  subjective metrics that may affect cloud market performance via literature survey. We used a popular sentiment analysis technique to process customer reviews collected from  social media. Cloud market revenue growth was selected as an indicator of cloud market performance. We considered the revenue growth of Amazon Web Services as the stakeholder of our experiments. Three machine learning models were selected: linear regression, artificial neural network, and support vector machine. These models were compared with a time series prediction model. We found that the  set of subjective metrics is able to improve the prediction performance for all the models. The support vector machine showed the best prediction results compared 
%10
 to the other models. 

\end{abstract}
\section{Introduction}

%\vspace*{-4mm}
Cloud computing is increasingly becoming the technology of choice as the next-generation platform for most organization \cite{sajib2018}\cite{SUN2019749}. Cloud computing services are provided by vendors under the following categories: Software-as-a-Service (SaaS), Platform-as-a-Service (PaaS), and Infrastructure-as-a-Service (IaaS) are available in the cloud market. Different organizations such as banks, universities, etc., are the main consumers of cloud services.   

The cloud market is divided into two types of participants: those who use cloud services - \textit{cloud customer}, and those who need to invest in the cloud market - \textit{cloud investors/providers}.  \textit{Both} the parties are interested in cloud service performance in the market. Sufficient knowledge of the cloud market allows the market participants to be aware of opportunities, advantages, and risks associated with it. This can include knowledge of how the market is performing in terms of market competition, market growth, customer satisfaction, etc. \cite{mauboussin2012true}, which can all help in making informed decisions regarding the market. In fact, a comprehensive analysis of the market is of value to the cloud customer due to the long-term commitment involved. Such customers could be those who make IT decisions after determining the influence of emerging technologies.  They must understand how the market performs in terms of customer satisfaction, service quality, and many other factors before deciding to participate in the market. Investors also require information on market performance before making any decision to invest in stocks. In fact, many industry reports have been prepared by different organisations in relation to all the participants \cite{USDEP}\cite{451Research}. Most of these reports are concerned with the evaluation of cloud market size and growth as well as numerous other concerns. These reports tend to be produced after thorough monitoring of the market or after accessing to cloud providers’ performance/sales.

The performance information provided by cloud market reports are based on two types of metrics: \textit{objective metrics} and \textit{subjective metrics}. The objective metrics are based on actual facts without any individual bias. Examples of such metrics include market size and growth. According to 451Research \cite{451Research}, the collection of such data is frequently implemented through extensive market monitoring. Meanwhile, the subjective metrics are based on personal perceptions, opinions or feelings, an example being the information regarding the barriers to cloud adoption \cite{STATE}. Subjective measures are frequently obtained using industry surveys \cite{RightScale}\cite{USDEP}. Industries utilise either objective and subjective measures or a combination of both to produce industry reports. However, it is a challenge to apply objective metrics \cite{vij2016subjective} due to the information sensitivity, while it can be \textit{expensive}, \textit{time consuming} and \textit{less timely} to obtain the information required for objective performance. On the other hand, the gathering of subjective market information can be more practical, since the sharing of subjective evaluation information is preferred over objective evaluation information sharing \cite{vij2016subjective}. Market participants often tend to share similar information on \textit{social media}.

Numerous individuals are becoming increasingly involved in sharing their opinions on a variety of subjects in social media platforms \cite{9001217}\cite{8029820}. It has become increasingly popular in recent years \cite{nepal2015trusting}. Consumers express their complaints or their satisfaction with certain brands and services. It has been reported that one in three social media users prefer to use this outlet rather than contact the company by phone. Moreover, 50\% of social media users share their complaints/concerns about a service/brand at least once a month \cite{STATE}. Hence, social media can be viewed as a \textit{free and up-to-date} source of personal wisdom \cite{yu2012survey}.

Purchase decisions can be affected by consumer's perceptions on services/goods. Thus, subjective measures can be used to predict the future behaviour of the participants. With this subjective measure evaluation of market performance, a better comprehension of objectively measured results can be gained. Any given customer review can include perceptions related to various areas of a product or service experience \cite{goh2013social}. In turn, these perceptions will influence the readers’ perceptions and, subsequently, their purchase decisions \cite{mudambi2010research}. Here, perceptions related to risks, barriers and benefits can be particularly influential on a consumer’s purchasing intention \cite{ruiz2009drivers}. Future behaviour can be predicted based on this intention to purchase \cite{shahabuddin2009forecasting}.

According to Jayaratna et al. \cite{Jayaratna}, subjective measures are applicable in evaluating cloud market performance.  They identified unique subjective metrics that can assess cloud market performance. However, the subjective information is limited to \textit{product/service quality}. In addition, attention has not been given to \textit{use machine learning techniques} for predicting economical/financial performance of cloud services in the cloud market. 

 We aim to \textit{explore a set of more comprehensive subjective metrics} that may affect cloud market performance and \textit{analyse the effectiveness of machine learning techniques} in predicting cloud market performance. A comprehensive set of subjective cloud service evaluation metrics were collected via literature survey. The values of subjective metrics were collected from social media and quantified using sentiment analysis. We applied three mainstream machine learning prediction techniques: \textit{linear regression}, \textit{artificial neural network}, and \textit{support vector machine} to ascertain which model predicts cloud market performance most accurately. We considered \textit{cloud market revenue growth} as the indicator of cloud market performance. This paper makes the following contributions:
\begin{enumerate}
\item We identified a set of more comprehensive  subjective metrics that may influence the cloud market performance.  
\item We explored an effective machine learning model to predict the cloud market performance based on the identified subjective metrics. 

\end{enumerate} 
% You must have at least 2 lines in the paragraph with the drop letter
% (should never be an issue)

The paper is structured as follows: Section 2 describes our solution. Section 3 depicts the results of our experiment. Section 4 concludes the paper.

\section{Methodology}
%\vspace*{-2mm}
We used two sets of metrics (subjective and objective) to evaluate cloud market performance. Cloud market revenue was considered as the objective metrics because it is an important factor in market performance evaluation \cite{mauboussin2012true}. Cloud user's perception on different cloud aspects was used as the subjective metrics. We explored an effective machine learning approach that  predicts cloud market revenue growth based on the cloud user's perceptions on different cloud aspects. The main objective of the proposed approach is to predict the cloud market performance based on a comprehensive set of subjective metrics.
%\vspace*{-4mm}
\subsection{Cloud Service Evaluation Subjective Metrics  Identification}
%\vspace*{-2.5mm}
Subjective metrics of customer perception have shown effectiveness in predicting cloud market performance over the objective metrics \cite{Jayaratna}. Therefore, in this research, we aim to explore a comprehensive set of subjective metrics that may influence cloud market performance.

Certain drivers and boundaries drive consumers to/from cloud services according to the surveys conducted by RightScale \cite{RightScale} and the Australian Bureau of Statistics \cite{ABS}. We can thus assume that drivers and boundaries are important for cloud consumers. The knowledge of the benefits and challenges are likely to influence a customer’s decision regarding the purchase of cloud services. Therefore, the benefits and challenges of cloud services will act as drivers of cloud market revenue. Jayaratna et al. \cite{Jayaratna} introduced 13 drivers and boundaries related to product/service quality.

Based on the above, we combined Jayaratna et al. \cite{Jayaratna}'s 13 drivers and boundaries with three more drivers: after-sales experience, market responsiveness and marketing execution  from Gartner \cite{gartner}, as a result of our literature survey. Table \ref{aspects} presents the combined drivers and boundaries used in this paper. We focused on consumer’s perception on same cloud aspects. In this paper, we aim to predict cloud market performance based on the 16 cloud aspects.
%\vspace*{-\baselineskip}
\begin{table}[]
\centering
\caption{Subjective aspects affecting cloud purchase decisions}
\label{aspects}
\resizebox{\textwidth}{!}{%
\begin{tabular}{|l|L{13cm}|}
\hline
\textbf{Cloud Aspects} & \textbf{Description}  \\ \hline
Greater scalability  & Flexible to either up-scale or down-scale   \\ \hline
 Faster access to infrastructure  & Easy access to infrastructure without having to purchase them    \\ \hline
  Managing multiple services  & Overheads and difficulty in managing multiple services  \\ \hline
   Security concerns   & Concerns over data breaches, privacy, access
control, etc.  \\ \hline
 Cost savings  & Cost saved by transfer to cloud infrastructure  \\ \hline

 Higher availability  & High availability of cloud services  \\  \hline
 Lack of control & Uncertainty of data location. Uncertainty
regarding legal issues, and dispute resolution   \\ \hline
 Higher performance  & Higher performance of cloud compared to
on-premise infrastructure  \\ \hline
 Lack of expertise/resources   & Lack of specialised people or sufficient resources for managing cloud services  \\ \hline
 
 IT staff efficiency   & Increase of productivity  \\ \hline
 Provider lock-in   & Difficulties with changing cloud computing service provider  \\ \hline
 Business continuity  & Ability to continually operate even through disasters  \\ \hline
 Move from CapEx to OpEx   & Changing from capital expenditure to operating expense  \\ \hline

 \textit{After-sales experience}   & Ability to provide acceptable customer services  \\ \hline
 \textit{Market responsiveness}   & Ability to enhance the products’ after sales   \\ \hline
 \textit{Marketing execution}   & Ability to deliver the product that the
customer expected \\ \hline
\end{tabular}%
}
\end{table}
%\vspace*{-9mm}
\subsection{Data Collection}
We collected two types of data: objective data and subjective data. The Amazon web service (AWS) quarterly revenue was considered as objective. The subjective data was collected in terms of AWS customer reviews.\\
\textbf{(1) Objective Data:\\}
IaaS is defined as the capability to provide computer resources, storage, and other fundamental computing resources to the customer. Accordingly, we consider the revenue generated by offering these types of cloud services. Amazon is considered as the leading cloud provider because it holds the largest market share in the IaaS segment. Therefore, we consider Amazon revenue as a cloud market performance. AWS revenue is reported for the whole segment (IaaS, PaaS, etc). In this paper, the AWS quarterly revenue growth \cite{AWS} from the 4th quarter of 2015 to the 4th quarter of 2018 was considered as an indicator of AWS performance in the market. Revenue growth was the increase in a company’s revenue from one period to the next. Revenue growth identifies over time trends in business. The AWS was calculated based on the following Equation \cite{Jayaratna}:
\begin{equation} \label{eq:growth}
AWS QRG_y = \frac{Revenue_q - Revenue_{q-1}}{Revenue_{q-1}}
\end{equation}
where $AWS QRG_y$ is the AWS quarterly revenue growth of quarter q, $Revenue_q$ is the revenue made in q, and $Revenue_{q-1}$ is the revenue made in the quarter before q.\\
\textbf{(2) Subjective Data:\\}
Social media is a rich source of customer opinion. Obvious differences can be seen among the customer perceptions collected from social media websites such as Twitter and Facebook and from customer review websites. However, social media websites also contain references to cloud services that are not based on experiences, such as references related to discussions or news on topics about newly-released features. In contrast, cloud customers’ review websites are exclusively focused on the customer experience. These review sites contain comprehensive information on users’ experiences. Moreover, any user can post comments about a cloud service/provider on social media, and the platforms do not verify whether a post is from a genuine cloud service customer. However, most of the popular review sites do go through this process before posting the feedback. Therefore, we can conclude that customer review websites provide more valuable information on cloud services than popular social media websites.

We examined six cloud social media, which are the most popular of those currently active. These are `G2 Crowd'\footnote{{https://www.g2crowd.com}}, `Trust Radius'\footnote{{https://www.trustradius.com}}, `Clutch'\footnote{{https://clutch.co/}}, `Gartner Peer Insights'\footnote{{https://www.gartner.com/reviews/home}}, `WhoIsHostingThis'\footnote{{https://www.whoishostingthis.com}} and `Spiceworks'\footnote{{https://www.spiceworks.com}}, all of which guarantee the genuineness of the reviews are based on the genuine customer experience. These sources contain reviews of a variety of cloud providers, including Amazon, Google, and Microsoft. The reviews are written by customers from different domains (i.e. education, health, insurance) and organisations of different scale (small, mid-sized and enterprise). We assumed that the customers’ reviews can be considered as a representative of the overall market customers because the number of customer reviews on these websites exceeds 1000. 
%\vspace*{-2mm}
\subsection{Perception Analysis on Cloud Aspects}
%\vspace*{-2mm}
VADER sentiment analysis technique \cite{hutto2014vader} is a well-acknowledged sentiment detection tool. VADER takes into account
punctuation, capitalization, degree modifiers, and the use of contrastive conjunction. In this paper, we used VADER sentiment analysis to process the user reviews. Sentiment analysis was performed to find customers' perception on each cloud aspect separately. First, we built a vocabulary that contains the most frequent words to refer to each specific aspect. We used NVivo software\footnote{{https://www.qsrinternational.com/nvivo/home}}, which has been designed for qualitative and mixed-methods research, to build the vocabulary. We fed the customer reviews of each quarter to the NVivo software. Then, we found the most frequent words that related to the cloud aspects. For example, secure, secured, securely, security were repeated more than 40 times in 2016 4th quarter reviews. Therefore, we assumed that those words will fit into the security cloud aspect.

Next, the sentiments of the AWS user reviews of each quarter were analysed using VADER. VADER is based on lexicons of sentiment related-words. Every word in the lexicon is rated as to whether it is positive or negative. VADER hired many domain experts to manually rate the words. Each word in the lexicon is rated between -4 (extremely negative) and +4 (extremely positive). VADER checks a piece of text to see if any of the words in the text are present in the lexicon.  VADER produces four sentiment scores: positive, neutral, negative, and compound. The positive, neutral, and negative scores are ratios for the proportions of text that falls into those categories. The compound score is a metric that calculates the sum of all the lexicon ratings\footnote{{http://comp.social.gatech.edu/papers/}} which have been normalized between -1 (most extreme negative) and +1 (extremely positive). Finally, we considered the compound sentiment results as the overall perception of a particular cloud aspect. The customers' perception of cloud aspect A in quarter Q is defined as follows:

\begin{equation}\label{SENTEQ}
Perception_{A,Q}= \frac{Compound_{A,Q}}{Reviews_{A,Q}}
\end{equation}  
where $Compound_{A,Q}$ is the compound sentiment value, and $Reviews_{A,Q}$ is the number of reviews referring to aspect $A$ in quarter $Q$.
%\vspace*{-\baselineskip}
\subsection{Sensitivity Analysis}
We designed the following machine learning models to evaluate cloud market performance based on the subjective metrics.\\
\textbf{(1) Linear Regression Analysis (LR):\\}
Linear regression is a linear approach used to ascertain the relationship between a dependent variable and one or more independent variables. In our approach, the independent variables are sets of cloud aspects, while the dependent variable is the AWS revenue growth. The linear regression has the following Equation: 
\begin{equation}
Y = a + bX_1 + cX_2 + ... + mX_n
\end{equation}
where $Y$ is used to represent the AWS revenue growth, and $X_1$, $X_2$, ... $X_n$ represent the user perception of each cloud aspect.\\
\textbf{(2) Artificial Neural Network (ANN):\\}
ANN models are mathematical models inspired by the functioning of the nervous system. ANNs are based on learning which is a characteristic of adaptive systems which are capable of improving their performance on a problem as a function of previous experience \cite{luk2000study}. In this paper, we used a Multlayer Perceptron. The Levenberg-Marquardt backpropagation algorithm the most widely-used training algorithm for time series prediction \cite{more1978levenberg} was used to train the network. The input activation function was obtained using the following Equation \cite{more1978levenberg}:
\begin{equation}
   s = \sum w_{ij}x_j + b 
\end{equation}
where $x_i$ is the user perception of each cloud aspect, $w_j$ is the weight, and b is the bias.
 
The output neuron uses the sigmoid activation function obtained with the following Equation \cite{more1978levenberg}:
\begin{equation}
f(s) = \frac{1}{1+e^{-s}}
\end{equation}
where $s$ is the sum of products (sop) between each user perception on each cloud aspect and its corresponding weight. $s$ is an input activation function.

The error function ascertains how close the predicted output is from the target output. The error function was obtained using the following Equation \cite{more1978levenberg}: 
\begin{equation}
E= \frac{1}{2} \sum_{i=1}^N (O_i-T_i)
\end{equation}
where $N$ is the total number of output data, $O_i$ is the actual AWS revenue growth of $i^{th}$ data and $T_i$ is the predicted AWS revenue growth of $i^{th}$ data.\\
\textbf{(3) Support Vector Machine (SVM):\\}
The nu-SVR model, which is applicable for modelling continuous time series, was chosen. The nu-SVR model was found to be reliable and robust, even for models based on small training samples or data burdened by noise based on experience from applications in a different area \cite{zhu2007performance}. In our approach, we used the nu-SVR model with a radial basis function kernel (RBF) and which takes the user perception of cloud services and AWS revenue growth as the target. The RBF kernel can be defined as follows \cite{zhu2007performance}:

\begin{equation}
\exp ( \gamma\times\sum \mid u-v \mid^2 )
\end{equation}
where $\gamma$ is the parameter, $u$ is the actual AWS revenue growth, and $v$ is the predicted AWS revenue growth.
\section{Experiments}
%\vspace*{-3mm}
A set of experiments was conducted to evaluate the efficiency of the proposed prediction models. First, we describe the dataset and how is it obtained. We then  compare the parameter of each model. Next, we  compare the prediction performance between four predictions models. We ran two experiments in order to analyse the impact of our new subjective metrics.  First, we used the user perceptions on the 13 cloud aspects identified in Jayaratna et al.'s \cite{Jayaratna}. Then, we used the user perceptions on the 16 cloud aspects in order to find the impact of the new subjective metrics. We evaluated all the proposed models in terms of Mean Squared Error (MSE), Root Mean Squared Error (RMSE) and Theil’s U-statistics.
%\vspace*{-4mm}
\subsection{Datasets}
\begin{figure}[]
\centering
\includegraphics [width=0.8\linewidth]{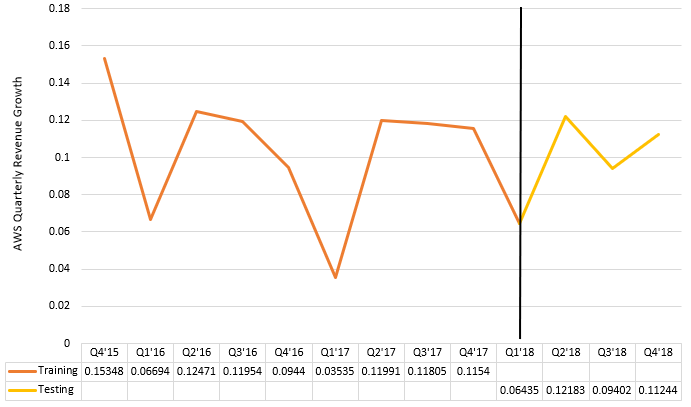}
\caption{AWS quarterly revenue from 2015 4th quarter to 2018 4th quarter}\label{fig:awsgrowth}
\end{figure}

%\vspace*{-2mm}
We divided the data set into the sets of training data and testing data according to an approximate ratio of 2:1. AWS quarterly revenue growth and user perceptions of each cloud aspect from the 4th quarter of 2015 to the 4th quarter of 2017 are the training data. The data from the 2018 1st quarter to 2018 4th quarter are the testing data. Figure \ref{fig:awsgrowth} presents the AWS quarterly revenue growth data. We collected approximately 1100 user reviews  for analysis.
%\vspace*{-\baselineskip}
%\vspace*{-1mm}
\subsection{Sentiment Analysis}
%\vspace*{-2mm}
First, we identified the most frequent words from the customer reviews of each quarter. For example, After-sales aspect has ten words which are customer service, satisfaction, good service, after-sales,  client service,  product service,  troubleshooting,  assistance,  customer care and  support. Next, we categorized the customer reviews based on the 16 cloud aspects that were introduced in the previous section. Next, the sentiment of the AWS user reviews were analysed. We processed each review to get the sentiment results. We then calculated the user perception using Equation \ref{SENTEQ}. Table \ref{sent} shows the sentiment results and the user perceptions for After-sales for 2016 4th quarter as an example.
% Please add the following required packages to your document preamble:
% \usepackage{graphicx}
%\vspace*{-5mm}
\begin{table}[]
\centering
\caption{Sentiment and user perceptions results for After-sales for 2016 Q4}
\label{sent}
\resizebox{\textwidth}{!}{%
\begin{tabular}{|l|l|l|l|l|l|l|l|l|l|l|l|l|}
\hline
\textbf{Review} & 1 &2  &3  &4  &5  &6 &7  &8  &9  &10  &11  & 12 \\ \hline
\textbf{Compound} &0.4215  &0.3612  &0.3182  &0.7092  &0.9196  &0.8876  &0.9681  &0.4091  &0.9702  &0.1027  &0.9715  &0.9829  \\ \hline
\textbf{User perception} & \multicolumn{12}{l|}{0.66848334} \\ \hline
\end{tabular}%
}
\end{table}
\vspace*{-10mm}
\subsection{Tuning Process for the Proposed Models}
We tested four models in the experiment, including an objective model (ARIMA) based on previous quarterly revenue growth and three subjective models based on previous quarterly revenue growth and customer perceptions. Here we introduced the optimal tuning result for each model.\\
\textbf{(1) ARIMA model:\\}
ARIMA model is a prediction model based on the objective metrics. Therefore, we considered the ARIMA model as our baseline. ARIMA stands for Autoregressive integrated moving average. ARIMA model is the most prominent methods in time series forecasting. ARIMA models have shown efficient capability to generate short-tram prediction \cite{tascikaraoglu2014review}.  In ARIMA model, the future value of a variable is a linear combination of past values and past errors, expressed as follows:
\begin{equation}
Y_t = \phi + \phi_1 Y_{t_{-_1}} + \phi_2 Y_{t_{-_2}} + ... + \phi_p Y_{t_{-_p}} +\varepsilon - \phi_1 \varepsilon_1{_{t_{-{_1}}}} - \phi_2 \varepsilon_2{_{t_{-{_2}}}} - ... - \phi_q \varepsilon_q{_{t_{-{_q}}}}
\end{equation}
where $Y_t$ is the AWS revenue growth, $\varepsilon$ is the random error at $t$, $\phi_i$ is the coefficients, $p$ and $q$ are integers that are often referred to as autoregressive and moving average, respectively. 

A standard notation is used of ARIMA(p,d,q). The autoregressive (p), integrated(d) and moving average (q) parameters have to be effectively determined in order to construct the best ARIMA model for AWS revenue growth prediction. We tested the model on different parameters of (p) and (q) (i.e., ARIMA (1,0,0), (1,1,0), (2,1,0), (0,1,1), (2,0,0) and (3,0,0)). We found that ARIMA (1,0,0), (2,1,0) and (2,0,0) achieve the best performance. All the experiments were conducted using Python.

%\begin{table}[]
%\centering
%\caption{Statistical results of different ARIMA parameters}
%\label{arimaPAR}
%\resizebox{0.8\textwidth}{!}{%
%\begin{tabular}{|l|l|l|l|l|l|l|}
%\hline
% \textbf{ARIMA Parameters}&(1,0,0)  &(1,1,0)  &(0,1,1)  & (2,1,0) & (2,0,0) & (3,0,0) \\ \hline
%\textbf{RMSE} & 0.024 & 0.032 & 0.027 & 0.024 & 0.024 & 0.032 \\ \hline
%\end{tabular}%
%}
%\end{table}
%\vspace*{-4mm}
\textbf{\\(2) Linear Regression Analysis (LR):\\}
At first, we tested the 13 cloud aspects as the independent variables to identify the prediction model. The prediction model based on the 13 cloud aspects can be represented in the Equation \ref{model13}. Then,  we tested the 16 cloud aspects as the independent variables to identify the prediction model. The prediction model can be represented in the Equation \ref{model16}.
\begin{multline}\label{model13}
RG= - 0.151*Per_{IT}-0.069*Per_{Perf} + 0.075*Per_{Cost} \\- 0.138*Per_{control} -0.113*Per_{Ex}-0.156*Per_{Provider}+0.042*Per_{Security} + 0.519
\end{multline}
%\vspace*{-10mm}
\begin{multline}\label{model16}
RG= -0.017*Per_{BC} + 0.054*Per_{Cost} - 0.065*Per_{control} -0.138*Per_{Ex}\\-0.132*Per_{Provider}-0.092*Per_{execution}-0.115*Per_{Security} + 0.515
\end{multline}
where $RG$ is the revenue growth, $Per_{BC}$ is the perception of business continuity, $Per_{Cost}$ is the perception of cost savings, $Per_{control}$ is the perception of lack of control, $Per_{Ex}$ is the perception of lack of expertise/resources, $Per_{Provider}$ is the perception of provider lock-in, $Per_{execution}$ is the perception of marketing execution, $Per_{Security}$ is the perception of security concerns, $Per_{IT}$ is the perception of IT staff efficiency  and $Per_{Perf}$ is the perception of higher performance.\\
\textbf{(3) Artificial Neural Network (ANN):\\}
We needed to establish the network parameters to build the neural network. The input data was randomly divided into training data (70\%), validation data (15\%) and testing data (15\%). The best validation performance for the 13 cloud aspects is obtained at epoch 2 and for the 16 cloud aspects is obtained at epoch 3. All experiments were conducted using the neural networks toolbox of MATLAB.\\
%\begin{figure}
%\centering
%  \mbox{\subfloat[]{\label{fig:ANNPP:1} \includegraphics[height=4.5cm]{nn_perf_13}}}
 % \mbox{\subfloat[]{\label{fig:ANNPP:2} \includegraphics[height=4.5cm]{nn_perf_16}}}
%  \caption{ANN Training Performance for (a) 13 cloud aspects, and (b) 16 cloud aspects  }
%\end{figure}
\textbf{(4) Support Vector Machine (SVM):\\}
Several levels of gamma constant of radial function was tested in the parameter settings experiments. 
We chose four different values for gamma 0.01, 0.1, 1, and 5  \cite{englucasstock}. We found that the best performance is obtained when the value of gamma is 5 for the both cloud aspect combinations. All experiments were performed on windows using Chang and Lin’s library developed for SVM implementations \cite{CC01a}.
\subsection{Experimental Result and Discussion}
\begin{figure}[]
\centering
\includegraphics[scale=0.45]{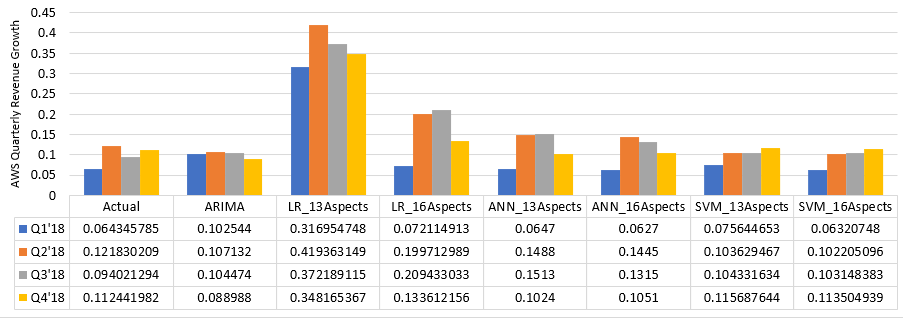}
\caption{Actual vs. Predicted }\label{model:1}
\end{figure}

%\vspace*{-2mm}
We predicted the AWS revenue growth form the 1st quarter of 2018 to the 4th quarter of 2018 using four different prediction models: ARIMA, LR, ANN, and SVM. Figure \ref{model:1} depicts the results of predicted revenue growth compared to the actual revenue growth. It can be seen that there is a significant gap between the actual and predicted revenue growth for LR with 13 and 16 cloud aspects prediction models. However, the other models were able to predict AWS revenue growth for all quarters.

%Figures \ref{rmse:2}, \ref{mse:2}, and \ref{u:2} shows the prediction performance results for each prediction models. Most of the prediction models except LR with 13 and 16 cloud aspects were able to predict accurately AWS revenue growth. SVM with 16 cloud aspects achieves superior performance in terms of RMSE (0.01085), MSE (0.000118), and Theil’s U-statistics (0.34369). It can be seen that the three new subjective metrics are improving both models. For the ANN prediction model,  It can be observed that there is a significant gap between the prediction result using 13 cloud aspects and 16 cloud aspects. SVM prediction model also performs better when we adding the new subjects metrics. Overall, adding the new subjective metrics (After-seals, Market responsiveness, Market execution) is improving the prediction performance. Our proposed models perform better than ARIMA model which is based on objective metrics. ANN and SVM were able to predict the AWS revenue growth for 2018 quarters with the lowest RMSE (0.0222, 0.01085), and MSE(0.000494, 0.00118) respectively. It can also be seen that SVM achieves the best prediction performance in terms of Theil’s U-statistics (0.34369).

All the experimental results prove the positive impact of the new set of subjective matrices and the prediction performance of the proposed models. All the models based on the 16 cloud factors outperform their counterparts based on the 13 cloud factors as described in Table \ref{per_allmodel}. We can confidently conclude that  \textit{After-seals}, \textit{Market responsiveness} and \textit{Market execution} improve the prediction performance. Additionally, the SVM prediction model with 16 cloud aspects produced the best results with the smallest error. The SVM prediction models outperformed ARIMA, ANN and LR models. The reason is that SVM implements the structural risk minimization principle, resulting in better generalization than the other techniques. In contrast, ANN needs a larger training set to perform more accurate predication, while ARIMA purely bases on the previous revenue data and disregards the impact of the subjective metrics on the avenue growth. LR preformed the worst due to the existence of nonlinear relationship between revenue growth and the subjective metrics. 
%\vspace*{-1mm}
%\vspace*{-6mm}
\begin{table}[]
\centering
\caption{Prediction performance of ARIMA, LR, ANN, and SVM}
\label{per_allmodel}
\resizebox{0.8\textwidth}{!}{%
\begin{tabular}{|l|l|l|l|}
\hline
\textbf{Prediction Model} &\textbf{MSE}  &\textbf{RMSE}  &\textbf{ Theil’s U-statistics} \\ \hline
 \textbf{ARIMA}&0.000583622  &0.024   &0.394573683  \\ \hline
 \textbf{LR - 13 Cloud Aspects}&0.071319997  &0.267058041 & 1.455854282 \\ \hline
 \textbf{LR - 16 Cloud Aspects}&0.004973533  &0.03526164  & 0.504616664  \\ \hline
 \textbf{ANN - 13 Cloud Aspects}&0.001027297  &0.0321  & 0.444061429  \\ \hline
 \textbf{ANN - 16 Cloud Aspects}&0.000493797  &0.0222  & 0.404017853  \\ \hline
 \textbf{SVM - 13 Cloud Aspects}&0.000143942  &0.011997583  & 0.354241995  \\ \hline
 \textbf{SVM - 16 Cloud Aspects}&\textbf{0.000117719}  &\textbf{0.010849839}  & \textbf{0.343687709} \\ \hline
\end{tabular}%
}
\end{table}
%\vspace*{-10mm}
\section{Conclusion}
We explored a machine learning based approach to predict the cloud market performance using social media subjective metrics. We identified a set of comprehensive subjective metrics (including After-seals, Market responsiveness, and Market execution). We processed the user reviews collected from social media using VADER sentiment analysis, a popular sentiment analysis technique on each subjective metric. We applied three mainstream machine learning: linear regression, artificial neural network, and support vector machine. Experimental results show that the new set of subjective metrics have a positive impact on predicting cloud market performance. The proposed model is able to predict cloud market performance for both cloud consumers and cloud investors/providers. In the future work, we plan to apply deep learning models to asses cloud market performance.
%\vspace*{-\baselineskip}
%\subsubsection*{Acknowledgment.}
%The  first  author  is  supported  by  a  scholarship  from Taibah University in Saudi Arabia.

%
% ---- Bibliography ----
%
% BibTeX users should specify bibliography style 'splncs04'.
% References will then be sorted and formatted in the correct style.
%
% \bibliographystyle{splncs04}
% \bibliography{mybibliography}
%

\bibliographystyle{splncs04}

\bibliography{Bib/main}
\end{document}